# Analysis of chromatic dispersion compensation and carrier phase recovery in long-haul optical transmission system influenced by equalization enhanced phase noise

Tianhua Xu[a,b,c*], Gunnar Jacobsen[b,d], Sergei Popov[b], Jie Li[d], Sergey Sergeyev[e], Ari T. Friberg[b,f], Tiegen Liu[c], Yimo Zhang[c]

[a]University College London, London, WC1E7JE, United Kingdom
[b]Royal Institute of Technology, Stockholm, SE-16440, Sweden
[c]Tianjin University, Tianjin, 300072, China
[d]Acreo Swedish ICT AB, Stockholm, SE-16440, Sweden
[e]Aston University, Birmingham, B47ET, United Kingdom
[f]University of Eastern Finland, Joensuu, FI-81010, Finland

* E-mail address: tianhua.xu@ucl.ac.uk, tianhuaxu@outlook.com

## ABSTRACT

The performance of long-haul coherent optical fiber transmission system is significantly affected by the equalization enhanced phase noise (EEPN), due to the interaction between the electronic dispersion compensation (EDC) and the laser phase noise. In this paper, we present a comprehensive study on different chromatic dispersion (CD) compensation and carrier phase recovery (CPR) approaches, in the *n*-level phase shift keying (*n*-PSK) and the *n*-level quadrature amplitude modulation (*n*-QAM) coherent optical transmission systems, considering the impacts of EEPN. Four CD compensation methods are considered: the time-domain equalization (TDE), the frequency-domain equalization (FDE), the least mean square (LMS) adaptive equalization are applied for EDC, and the dispersion compensating fiber (DCF) is employed for optical dispersion compensation (ODC). Meanwhile, three carrier phase recovery methods are also involved: a one-tap normalized least mean square (NLMS) algorithm, a block-wise average (BWA) algorithm, and a Viterbi-Viterbi (VV) algorithm. Numerical simulations have been carried out in a 28-Gbaud dual-polarization quadrature phase shift keying (DP-QPSK) coherent transmission system, and the results indicate that the origin of EEPN depends on the choice of chromatic dispersion compensation methods, and the effects of EEPN also behave moderately different in accordance to different carrier phase recovery scenarios.

## 1. Introduction

Long-haul high speed optical fiber communications pose strict requirements of their tolerance to the linear and the nonlinear channel distortions [1-3]. Coherent optical transmission employing digital signal processing (DSP) allows the compensation of system impairments, such as chromatic dispersion (CD), polarization mode dispersion (PMD), laser phase noise (PN), and fiber nonlinearities (FNLs), in the electrical domain [4-12]. Using powerful DSP algorithms, the equalization of fiber chromatic dispersion and the compensation of laser phase noise have been carried out effectively in coherent transmission systems according to the reported work [3-10]. The electronic dispersion compensation (EDC) can be implemented by using the digital filters in both the time domain and the frequency domain [4-6,13-16], and the carrier phase recovery (CPR) can be realized by using the feed-forward and the feed-back DSP algorithms [8-10,17-19]. However, in these conventional EDC and CPR algorithms, the analysis of the phase noise from the transmitter (Tx) and the local oscillator (LO) lasers was always lumped together, where the interplay between the dispersion compensating module (DCM) and the laser phase noise was not considered.

Recently, the interaction between the electronic dispersion equalization and the laser phase noise, which leads to an effect of equalization enhanced phase noise (EEPN), has attracted the research interests due to its significant impacts in long-haul optical transmission systems [20-35]. W. Shieh et al. have reported the theoretical assessment of the EEPN from the enhanced LO phase noise, and they also investigated the EEPN induced time jitter in coherent transmission systems [20-23]. C. Xie has studied the effects of EDC on both the LO phase noise to amplitude noise conversion and the fiber nonlinear interference [24,25]. I. Fatadin et al. have investigated the influence of EEPN in the $n$-level quadrature amplitude modulation ($n$-QAM) systems, such as the quadrature phase shift keying (QPSK), the 16-QAM and the 64-QAM coherent optical communication systems [26]. The impacts of EEPN have also been investigated in the optical orthogonal frequency division multiplexing (OFDM) transmission systems [28,29]. Meanwhile, some experimental studies regarding EEPN have also been carried out in the QPSK transmission systems [30,31]. In addition, some approaches have also been investigated to mitigate the EEPN, by using the traditional CPR algorithms [32,33], the differential phase estimation [34], the pre-compensation of chromatic dispersion [35], the digital coherence enhancement [27], the optical reference carrier [36,37], and the partially modulated optical carrier [38,39]. Among these methods, the digital coherence enhancement can offer an effective compensation of EEPN [27], while it requires a complicated hardware implementation to measure the LO laser phase fluctuation. The impact of EEPN scales with the increment of fiber length, laser linewidth, symbol rate and modulation format, and significantly degrades the performance of the long-haul high speed coherent optical communication systems [20-34]. The conventional analysis of CD compensation and carrier phase recovery, which only takes into account the intrinsic Tx and LO lasers phase noise, is not suitable any longer for the long-haul coherent transmission system with a considerable EEPN. Therefore, it is of importance to investigate in detail the performance of different chromatic dispersion compensation and carrier phase recovery approaches in the long-haul coherent optical communication systems, where the influence of EEPN is not negligible.

In this paper, built on our previous work where the performance of carrier phase recovery was studied in the transmission system using frequency-domain dispersion compensation [32], we present a comprehensive investigation on different chromatic dispersion compensation and carrier phase recovery methods in the $n$-level phase shift keying ($n$-PSK) and $n$-QAM coherent optical transmission systems considering the impacts of equalization enhanced phase noise. Four chromatic dispersion compensation methods are considered, including the time-domain equalization (TDE), the frequency-domain equalization (FDE), the least mean square (LMS) adaptive equalization for EDC, and the dispersion compensating fiber (DCF) for optical dispersion compensation (ODC). Three carrier phase recovery methods are applied for the laser phase noise compensation: a one-tap normalized least mean square (NLMS) algorithm, a block-wise average (BWA) algorithm, and a Viterbi-Viterbi (VV) algorithm. The origin and the impact of EEPN are analyzed and discussed in detail by using and comparing different chromatic dispersion compensation and carrier phase recovery approaches. Numerical simulations have been implemented in a 28-Gbaud non-return-to-zero dual-polarization QPSK (NRZ-DP-QPSK) coherent optical transmission system, based on the Virtual Photonics Instruments (VPI) and the Matlab software [40,41]. Simulation results indicate that the origin of EEPN depends on the choice of chromatic dispersion compensation methods, and the effects of EEPN behave moderately different in diverse carrier phase recovery approaches. In the transmission system using EDC, the performance of the system employing the TDE and the FDE dispersion equalization is significantly affected by the equalization enhanced LO phase noise (EELOPN). However, in the LMS adaptive dispersion equalization, the system performance is equally influenced by the equalization enhanced Tx phase noise (EETxPN) and the equalization enhanced LO phase noise. There is no EEPN in the transmission system using optical dispersion compensation. In the study of CPR approaches for mitigating EEPN, the one-tap NLMS algorithm gives a marginally worse (but still acceptable) performance than the block-wise average and the Viterbi-Viterbi approaches, when all the CPR methods are applied with an optimum operation. Meanwhile, the Viterbi-Viterbi algorithm only performs slightly better than the block-wise average algorithm, even though it requires more computational complexity. Our analysis and discussions are helpful and important for the practical design and application of the chromatic dispersion compensation and the carrier phase recovery in long-haul high speed coherent optical fiber transmission systems, where the EEPN cannot be neglected.

## 2. Equalization enhanced phase noise in coherent transmission system

The schematic of coherent optical fiber transmission system employing electronic CD equalization and carrier phase recovery is illustrated in Fig. 1.

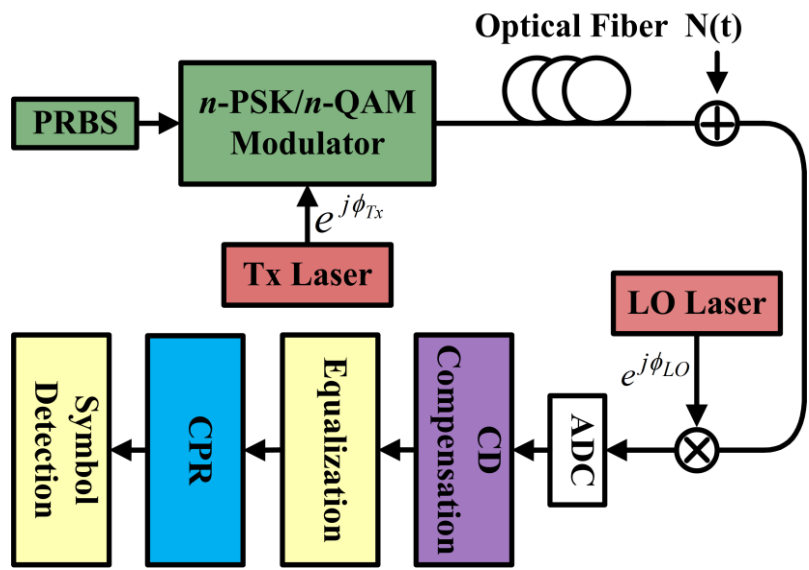

**Fig. 1.** Schematic of EEPN in coherent transmission system using EDC. N(t): additive white Gaussian noise (AWGN), ADC: analog-to-digital convertor.

As an example, we consider the use of a fixed TDE or FDE for the CD equalization [4,15,17,20]. In such cases, the phase noise from Tx laser passes through both the optical fiber and the EDC module, and the net experienced dispersion is close to zero. By contrast, the phase noise from LO laser only goes through the EDC module, which is heavily dispersed in the transmission system without using any optical dispersion compensation. Therefore, the LO phase noise will interplay with the EDC module, and the induced EEPN will significantly affect the performance of the long-haul high speed optical fiber networks. On the contrary, there is no EEPN phenomenon in the transmission systems using optical dispersion compensation, such as DCFs and chirped fiber Bragg gratings (FBGs), since both the Tx laser phase noise and the LO laser phase noise will experience a net dispersion of zero in such systems.

It has been reported that, the impact of EEPN scales linearly with the accumulated dispersion, the LO laser linewidth, and the symbol rate [17,20,34], and the variance of the additional noise due to EEPN can be expressed as following,

$$\sigma_{EEPN}^2 = \frac{\pi \lambda^2}{2c} \cdot \frac{D \cdot L \cdot \Delta f_{LO}}{T_S} \quad (1)$$

where $\lambda$ is the central wavelength of the optical carrier, $c$ is the light speed in vacuum, $D$ is the CD coefficient of fiber, $L$ is the fiber length, $\Delta f_{LO}$ is the 3-dB linewidth of LO laser, and $T_S$ is the symbol period of the coherent transmission system.

Consequently, the total phase noise variance in the coherent transmission system including the EEPN can be expressed as [17,37]:

$$\begin{aligned}\sigma^2 &= \sigma_{Tx}^2 + \sigma_{LO}^2 + \sigma_{EEPN}^2 + 2\rho \cdot \sigma_{LO} \sigma_{EEPN} \\ &\approx \sigma_{Tx}^2 + \sigma_{LO}^2 + \sigma_{EEPN}^2\end{aligned} \quad (2)$$

$$\sigma_{Tx}^2 = 2\pi \Delta f_{Tx} \cdot T_S \quad (3)$$

$$\sigma_{LO}^2 = 2\pi \Delta f_{LO} \cdot T_S \quad (4)$$

where $\sigma^2$ represents the total phase noise variance, $\sigma_{Tx}^2$ and $\sigma_{LO}^2$ are the intrinsic phase noise variance of the Tx and the LO lasers respectively, $\Delta f_{Tx}$ is the 3-dB linewidth of the Tx laser, and $\rho$ is the correlation coefficient between the EEPN and the intrinsic LO phase noise, which will be close to zero with the transmission distance exceeding the order of 80 km [17].

According to the definition of the intrinsic phase noise from the Tx and the LO lasers, a concept of effective linewidth $\Delta f_{Eff}$ can be employed to evaluate the total phase noise in the coherent transmission system considering EEPN [17,36], as following

$$\begin{aligned}\Delta f_{Eff} &= \frac{\sigma_{Tx}^2 + \sigma_{LO}^2 + \sigma_{EEPN}^2 + 2\rho \cdot \sigma_{LO} \sigma_{EEPN}}{2\pi T_S} \\ &\approx \frac{\sigma_{Tx}^2 + \sigma_{LO}^2 + \sigma_{EEPN}^2}{2\pi T_S}\end{aligned} \quad (5)$$

## 3. Chromatic dispersion compensation

### 3.1 Time-domain CD equalization

The time-domain CD equalization can be implemented by using the convolution between the digital filter and the received symbols. The tap weights $W_{TDE}$ in the time-domain CD equalizer can be expressed by the following equations [4,5],

$$W_{TDE}(k) = \sqrt{\frac{jcT^2}{D\lambda^2 L}} \exp\left(-j\frac{\pi cT^2}{D\lambda^2 L}k^2\right) \qquad -\left\lfloor\frac{N_{TDE}}{2}\right\rfloor \leq k \leq \left\lfloor\frac{N_{TDE}}{2}\right\rfloor \qquad (6)$$

$$N_{TDE} = 2 \times \left\lfloor\frac{|D|\lambda^2 L}{2cT^2}\right\rfloor + 1 \qquad (7)$$

where $T = T_S/2$ is the sampling period (for the DSP using 2 samples per symbol), $k$ represents the index of the received symbol, $N_{TDE}$ is the required number of taps for compensating the fiber dispersion, and $\lfloor t \rfloor$ denotes the nearest integer less than $t$.

### 3.2 Frequency-domain CD equalization

The frequency-domain CD equalization can be realized by using the overlap-save or the overlap-add fast Fourier transform (FFT) method [14,15], where the transfer function of the frequency domain equalizer is given by

$$G_{FDE}(L,\omega) = \exp\left(-j\frac{D\lambda^2 L}{4\pi c}\omega^2\right) \qquad (8)$$

The tap weights $W_{FDE}$ in the frequency domain equalizer can be expressed as

$$W_{FDE}(k) = \exp\left[-j\frac{D\lambda^2 L}{\pi c}\cdot\left(\frac{k}{N_{FFT}}\omega_N\right)^2\right] \qquad -\frac{N_{FFT}}{2} \leq k \leq \frac{N_{FFT}}{2}-1 \qquad (9)$$

where $N_{FFT}$ is the FFT-size of the frequency domain equalizer, and $\omega_N$ is the Nyquist angular frequency of the transmission system.

In this work, the FDE for dispersion compensation is implemented using the overlap-add FFT method, in which the minimum required zero-padding value is determined by the fiber dispersion to be compensated, and can be calculated using the following equation [15, 42]:

$$N_{FDE}^{zero-padding} = 2 \times \left\lceil\frac{\sqrt{\pi^2 c^2 T^4 + 4\lambda^4 D^2 L^2}}{\pi cT^2} + 1\right\rceil \qquad (10)$$

where $\lceil t \rceil$ refers to the nearest integer larger than $t$.

In our work, the FFT-size in the FDE for dispersion compensation is determined by the value of power of two closest to and larger than twice the minimum required zero-padding value, since the radix-2 FFT algorithm is applied. The FFT-size in FDE for compensating the dispersion of different fiber lengths is illustrated in Table 1.

### 3.3 Least mean square adaptive CD equalization

The LMS CD equalization is also implemented by using the convolution between the digital filter and the received symbols, while it requires an iterative and successive correction on the tap weights vector to achieve the minimum mean square error [16]. The tap weights vector $\vec{W}_{LMS}$ in the LMS equalization can be expressed as follows:

$$\vec{W}_{LMS}(k) = \vec{W}_{LMS}(k-1) + \mu_{LMS}\vec{x}(k-1)e^{*}_{LMS}(k-1) \tag{11}$$

$$e_{LMS}(k-1) = d_{LMS}(k-1) - \vec{W}^{H}_{LMS}(k-1)\vec{x}(k-1) \tag{12}$$

where $\vec{x}(k)$ is the received symbol vector, $\mu_{LMS}$ is the step size, $d_{LMS}(k)$ is the desired output symbol, $e_{LMS}(k)$ is the estimation error, $H$ is the Hermitian transform, and * is the conjugate operation.

## 4. Carrier phase recovery

### 4.1 One-tap NLMS carrier phase recovery

The one-tap NLMS algorithm can be employed for an effective phase estimation and carrier recovery [6,9], and the tap weight $W_{NLMS}$ in the one-tap NLMS CPR algorithm can be expressed as:

$$W_{NLMS}(k) = W_{NLMS}(k-1) + \frac{\mu_{NLMS}}{|x(k-1)|^2}x^{*}(k-1)e_{NLMS}(k-1) \tag{13}$$

$$e_{NLMS}(k-1) = d_{NLMS}(k-1) - W_{NLMS}(k-1)x(k-1) \tag{14}$$

where $x(k)$ is the input symbol, $\mu_{NLMS}$ is the step size parameter, $d_{NLMS}(k)$ is the desired output symbol, and $e_{NLMS}(k)$ is the estimation error between the output symbol and the desired symbol.

### 4.2 Block-wise average carrier phase recovery

The block-wise average algorithm calculate the *n*-th power of the input symbols in each processing unit (PU) to remove the phase modulation information, and the residual phase information are summed and averaged over the entire PU to minimize the impact of additive noise (the length of PU is called block size). Then the calculated phase is divided by *n*, to get the phase estimate for the entire PU [18]. Consequently, the estimated carrier phase for each PU in the BWA algorithm can be described as:

$$\phi_{BWA}(k) = \frac{1}{n}\arg\left\{\sum_{p=1+(m-1)\cdot N_{BWA}}^{m\cdot N_{BWA}} x^n(p)\right\}, \qquad m = \left\lceil \frac{k}{N_{BWA}} \right\rceil \qquad (15)$$

where $N_{BWA}$ is the block size in the BWA algorithm, $k$ is the index of the input symbol, and $\lceil t \rceil$ represents the nearest integer larger than $t$.

### 4.3 Viterbi-Viterbi carrier phase recovery

The Viterbi-Viterbi algorithm also calculates the *n*-th power of the input symbols in each PU to remove the phase modulation. Meanwhile, the calculated phase are also summed and averaged over the entire PU to minimize the effect of amplitude noise. However, the final estimated phase in the VV algorithm is only applied as the phase estimation for the central symbol in each PU [19]. The estimated carrier phase in the Viterbi-Viterbi algorithm can be described as:

$$\phi_{VV}(k) = \frac{1}{n}\arg\left\{\sum_{q=-(N_{VV}-1)/2}^{(N_{VV}-1)/2} x^n(k+q)\right\}, \qquad N_{VV}=1,3,5,7\ldots \qquad (16)$$

where $N_{VV}$ is the block size in the VV algorithm.

### 5. Simulation setup

As illustrated in Fig. 2, numerical simulations have been carried out in the 28-Gbaud NRZ-DP-QPSK coherent optical transmission system based on the VPI and the Matlab platforms [40,41]. The pseudo random bit sequence (PRBS) data with a pattern length of $2^{15}-1$ are modulated into two orthogonally polarized NRZ-QPSK optical signals by using the *n*-PSK (or *n*-QAM) modulators. Then the orthogonally polarized signals are fed into the fiber channel using a polarization beam combiner (PBC) to form the 28-Gbaud DP-QPSK signal. In the receiver end, the received optical signals are mixed with the LO laser and converted into electrical signals by using the photodiodes (PDs). The four electrical signals are processed by further using the 5-th order Bessel low-pass filters (LPFs) with a 3-dB bandwidth of ~19.6 GHz. Then they are sampled by the 8-bit analog-to-digital convertors (ADCs) at twice the symbol rate. The digitized signals are further processed by using the DSP modules to compensate the system impairments, and the bit-error-rate (BER) is measured based on a data sequence of $2^{18}$ bits. The central wavelength of the Tx and the LO lasers are both 1553.6 nm, and the standard single mode fibers (SSMFs) with a CD coefficient of 16 ps/nm/km are applied in all the simulation work. The fiber attenuation, the dispersion slope, the polarization mode dispersion (PMD) and the fiber nonlinear effects are neglected in the simulations.

The CD compensation module is realized by the TDE, the FDE, and the LMS adaptive equalization, and the carrier phase recovery is implemented by using the one-tap NLMS algorithm, the BWA algorithm, and the VV algorithm, respectively.

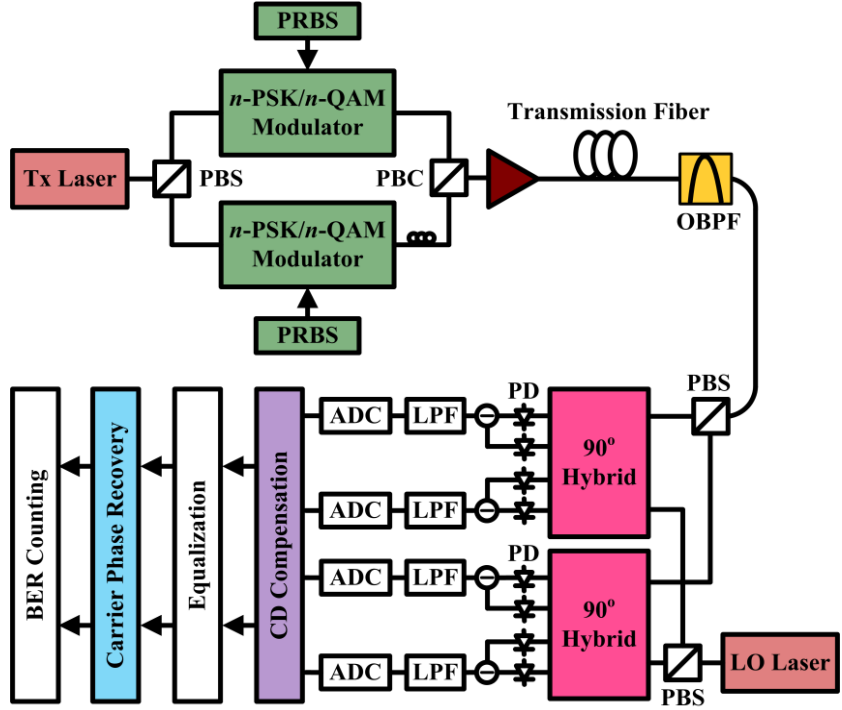

**Fig. 2.** Schematic of 28-Gbaud DP-QPSK coherent optical transmission system. PBS: polarization beam splitter, OBPF: optical band-pass filter.

The fiber lengths of 20 km, 40 km, 100 km, 400 km, 600 km, 1000 km and 2000 km have been employed in the simulations. The number of taps in TDE and the FFT-size in FDE for chromatic dispersion compensation are shown in Table 1.

**Table 1. Number of taps in TDE and FFT-size in FDE for different fiber lengths**

| Fiber length | Number of taps in TDE | FFT-size in FDE |
|---|---|---|
| 20 km | 9 | 16 |
| 40 km | 17 | 32 |
| 100 km | 41 | 64 |
| 400 km | 161 | 256 |
| 600 km | 243 | 512 |
| 1000 km | 403 | 1024 |
| 2000 km | 807 | 2048 |

## 6. Simulation results and analysis

### 6.1 Influence of chromatic dispersion compensation on EEPN

To focus on the phase noise enhancement effects due to different dispersion compensation approaches, the same carrier phase recovery method (the one-tap NLMS algorithm, which can be applied to any modulation format with a constant complexity [9,43]) is applied in the transmission systems using different CD compensation scenarios. Figure 3 shows the convergence of the tap weight in the one-tap NLMS algorithm, when different values of EEPN are considered. The linewidths of both the Tx and the LO lasers are 100 kHz in Fig. 3(a), and the linewidths of both the Tx and the LO lasers are 10 MHz in Fig. 3(b). In both cases, the transmission fiber length is 1000 km, and the FDE is employed for CD compensation. Thus we have an effective linewidth of ~2.7 MHz in Fig. 3(a), and an effective linewidth of ~270 MHz in Fig. 3(b), according to the definition in Eq. (5). It can be seen that the tap weight in Fig. 3(a) (less EEPN) converges a little faster and has a slightly smaller fluctuation (after converging, the fluctuation is ~10% of tap weight magnitude in Fig.3 (a) and is ~16% of tap weight magnitude in Fig.3 (b)) than the tap weight in Fig. 3(b) (larger EEPN), while in both cases the tap weight shows a good convergence (less than 400 iterations). Therefore, the one-tap NLMS algorithm can work well for tracking the phase change in the carrier phase estimation (CPE) in the transmission systems within a large range of EEPN (or effective linewidth) variation.

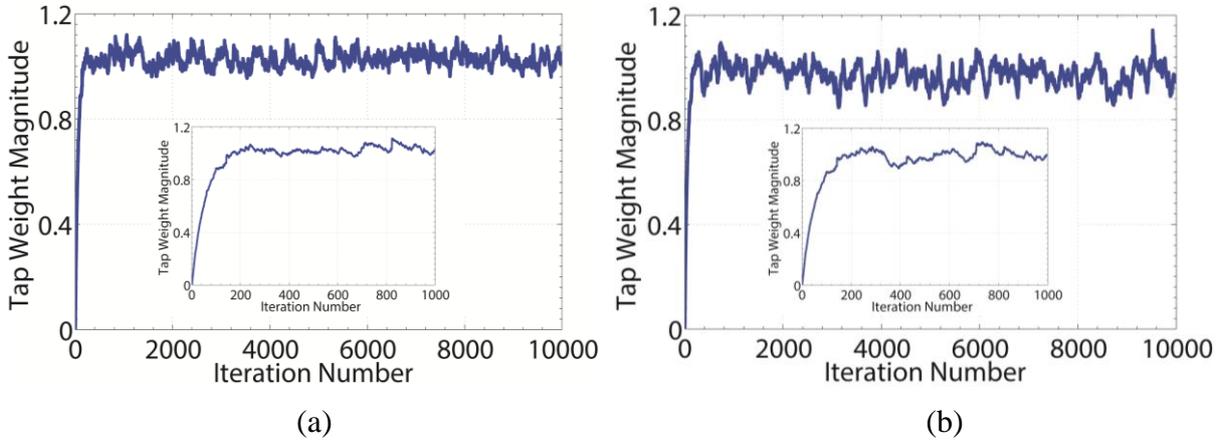

**Fig. 3.** Tap weight convergence in the one-tap NLMS CPR algorithm. (a) both Tx laser and LO laser linewidths are 100 kHz ($\Delta f_{Eff} \approx 2.7$ MHz), (b) both Tx laser and LO laser linewidths are 10MHz ($\Delta f_{Eff} \approx 270$ MHz).

In the following section, we will investigate the numerical simulations using the TDE, the FDE, the LMS adaptive equalization, and the ODC for chromatic dispersion compensation in long-haul transmission systems. In these simulations (Fig. 4, Fig. 5, Fig. 7, and Fig. 9), fiber lengths of 400 km and 2000 km are employed respectively, and different combinations of the Tx laser and the LO laser linewidths with a constant summation (4 MHz in Fig. 4, Fig. 5, Fig. 9, and 1 MHz in Fig. 7) are applied to investigate the origin and the effects of EEPN.

The bit-error-rate (BER) performance in the 28-Gbaud DP-QPSK coherent transmission system using the TDE for dispersion compensation is shown in Fig. 4, where an EEPN originating from the LO laser phase noise can be observed. From Fig. 4(a) to Fig. 4(c), we can see that the performance of the system degrades with the increment of the LO laser linewidth, when the one-tap NLMS carrier phase estimation is applied. In Fig. 4(b) and Fig. 4(c) (non-zero LO laser linewidth), the BER performance also degrades with increment of the transmission distance. Compared to the back-to-back (BtB) case, the EEPN induced optical signal-to noise ratio (OSNR) penalty in CPR scales with the LO laser linewidth and the accumulated fiber dispersion. Figure 5 shows the BER performance in the same 28-Gbaud DP-QPSK transmission system using the FDE for CD compensation, where a similar behavior of EEPN can be found as in Fig. 4. This indicates that the EEPN has the same origin and effects in the coherent transmission system using the TDE and the FDE for dispersion compensation.

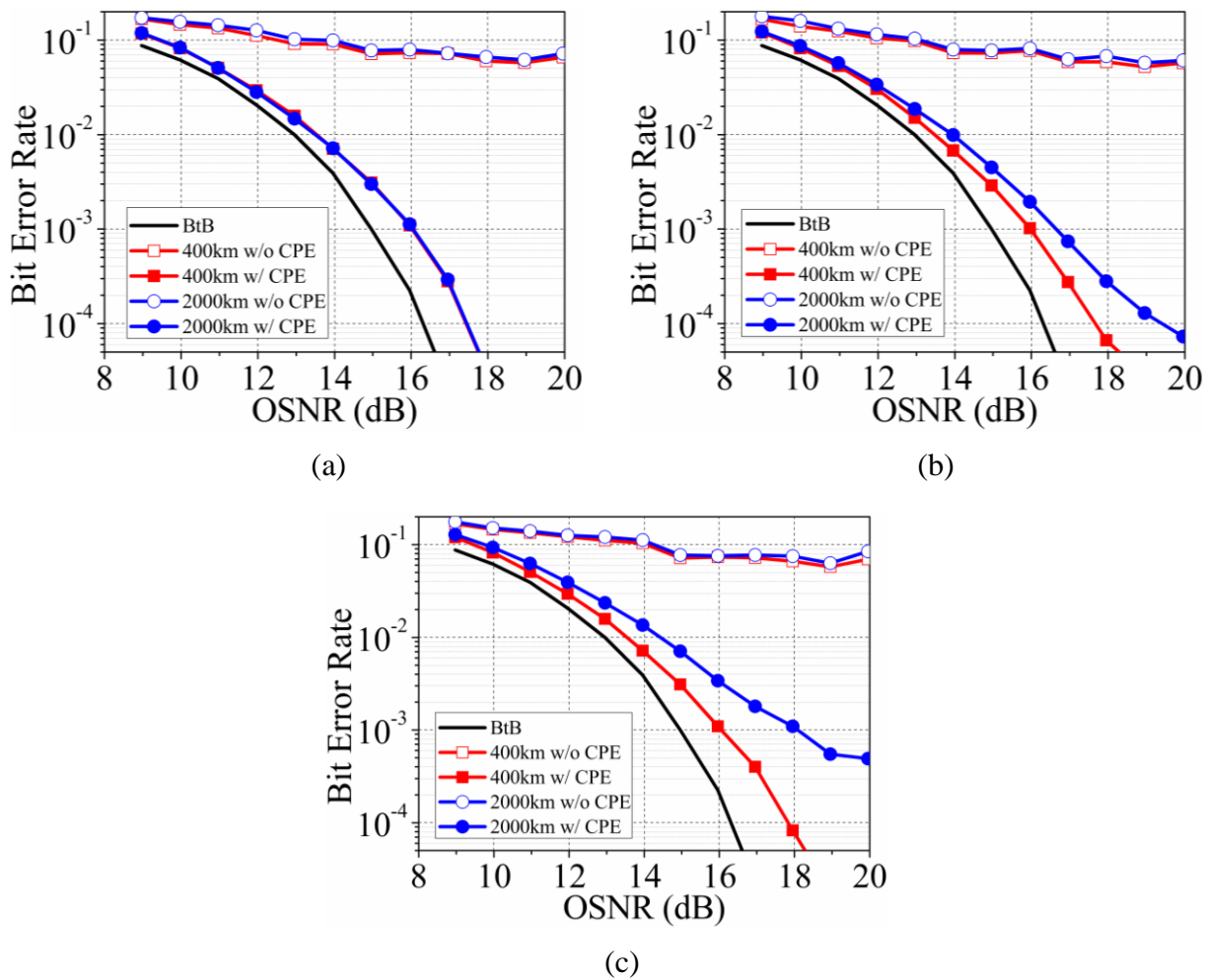

**Fig. 4.** Influence of EEPN in the DP-QPSK transmission system with different fiber lengths using TDE based dispersion compensation, (a) Tx laser linewidth is 4 MHz and LO laser linewidth is 0 Hz, (b) both Tx laser and LO laser linewidths are 2 MHz, (c) Tx laser linewidth is 0 Hz and LO laser linewidth is 4 MHz. w/o: without, w/: with.

Then the LMS adaptive equalization is also investigated for CD compensation. Figure 6 shows the BER performance of the 28-Gbaud DP-QPSK transmission system using LMS equalization for dispersion compensation, while no CPR is applied. In Fig. 6, we can find that the LMS algorithm can perform to some extent to compensate the carrier phase noise as well. Figure 6(a) shows the performance of a 20 km fiber transmission system with different laser linewidths (both Tx laser and LO laser linewidths are equal to the indicated value), and Fig. 6(b) shows the BER performance in a transmission system with different fiber lengths (both the Tx laser and the LO laser linewidths are 500 kHz). It can be found that the LMS algorithm can compensate the fiber dispersion and the laser phase noise simultaneously, and the performance of LMS equalization will degrade due to a more severe EEPN with the increment of the transmission distance and the laser linewidth.

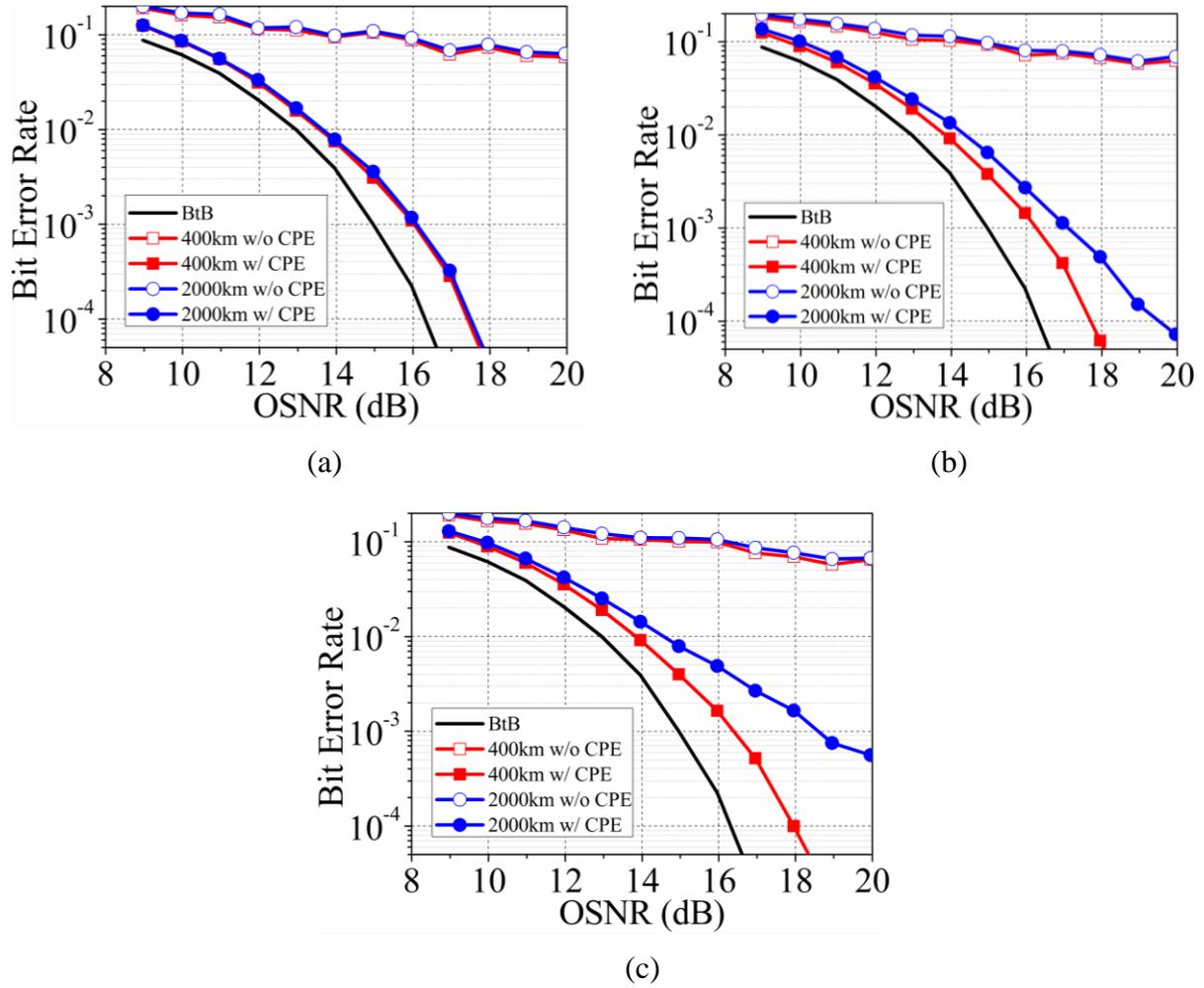

**Fig. 5.** Influence of EEPN in the DP-QPSK transmission system with different fiber lengths using FDE based dispersion compensation, (a) Tx laser linewidth is 4 MHz and LO laser linewidth is 0 Hz, (b) both Tx laser and LO laser linewidths are 2 MHz, (c) Tx laser linewidth is 0 Hz and LO laser linewidth is 4 MHz. w/o: without, w/: with.

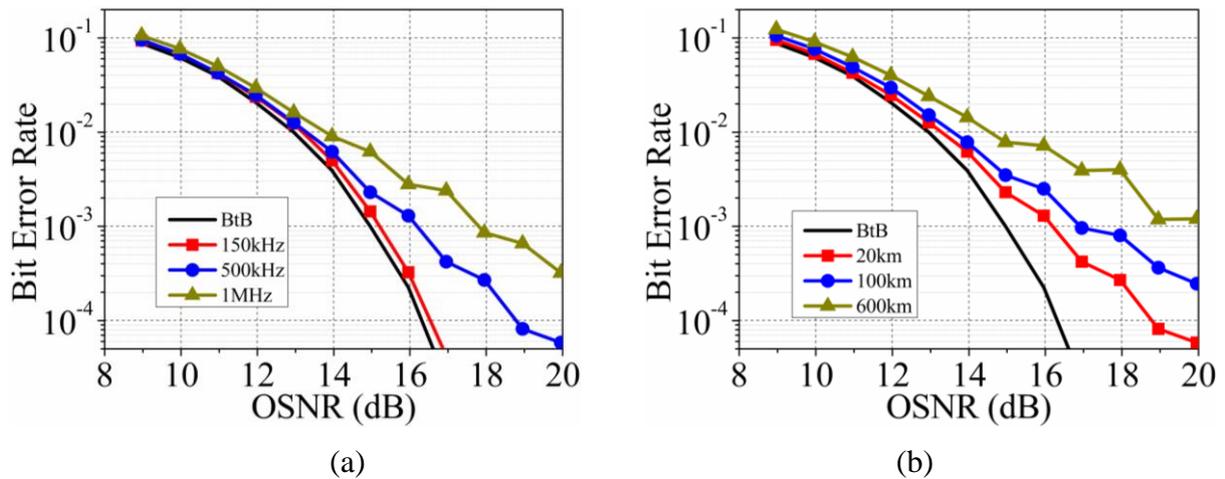

**Fig. 6.** Performance of 28-Gbaud DP-QPSK transmission system using LMS dispersion compensation (no CPR applied). (a) 20 km fiber with different laser linewidths, (b) different fiber lengths with both Tx and LO lasers linewidths of 500 kHz.

Under different Tx laser and LO laser linewidth distributions, the BER performance of the 28-Gbaud DP-QPSK transmission system employing the LMS dispersion compensation is shown in Fig. 7, where the one-tap NLMS algorithm is applied for the phase noise compensation. The simulation results are obtained under different combinations of the Tx laser and the LO laser linewidths with a constant summation (1 MHz). It can be found that the EEPN in the LMS equalization arises from the interplay between the dispersion equalization and the phase noise, while both the Tx and the LO phase noise are equally involved. Due to EEPN, the performance of LMS dispersion compensation degrades with the increment of the fiber length and the laser linewidth. The LMS dispersion equalization shows almost the same behavior in Fig. 7(a), Fig. 7(b), and Fig. 7(c), meaning that the LMS dispersion equalization interplays with the phase noise from both the Tx and the LO lasers equally. Moreover, the LMS equalization is less tolerant against the EEPN than the TDE and the FDE dispersion compensation, since Fig. 7 (EEPN generated from a total laser linewidth of 1 MHz) gives a similar BER behavior as in Fig. 4(c) and Fig. 5(c) (EEPN generated from a LO laser linewidth of 4 MHz).

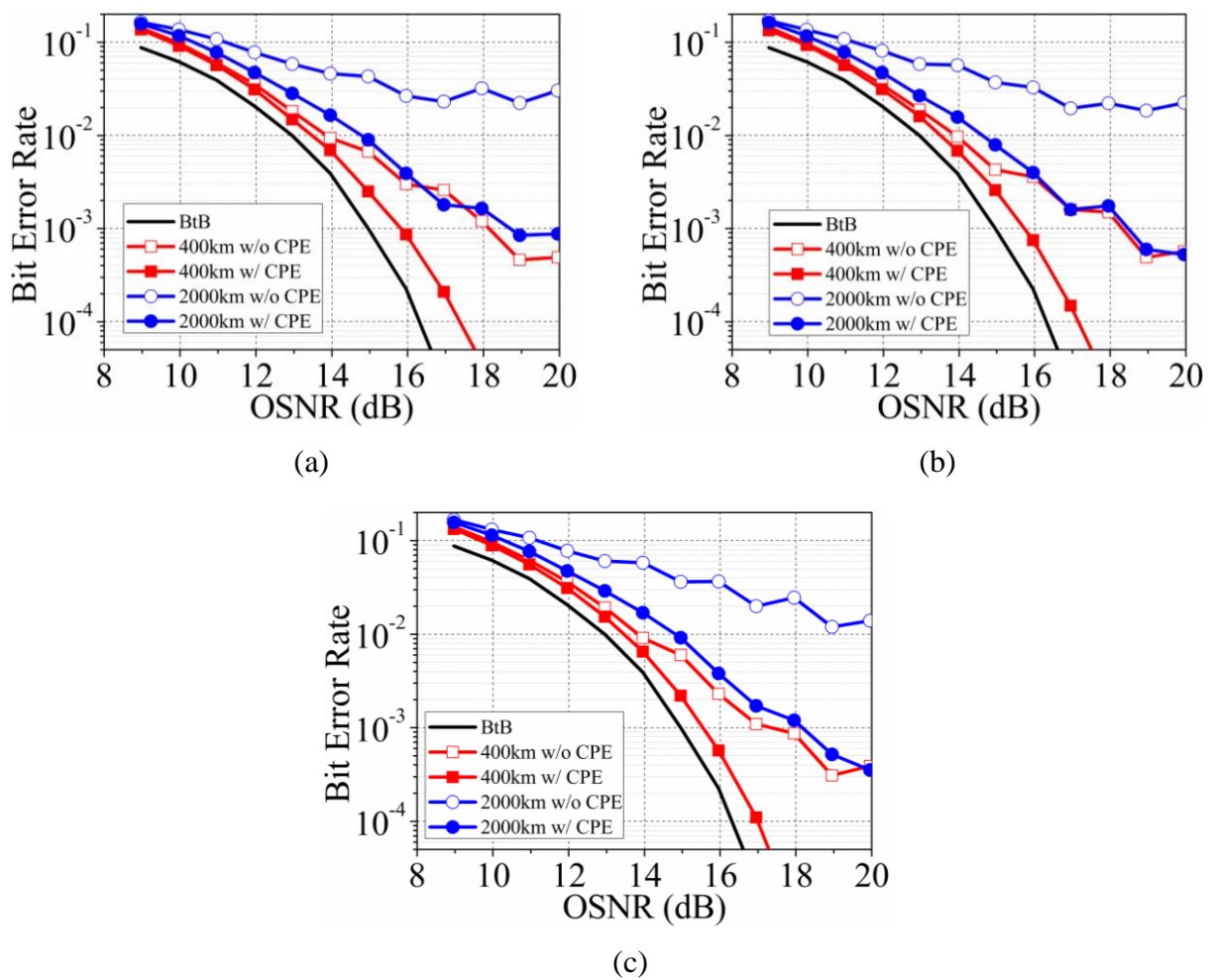

**Fig. 7.** Performance of 28-Gbaud DP-QPSK transmission system using LMS dispersion compensation with different fiber lengths, (a) Tx laser linewidth is 1 MHz and LO laser linewidth is 0 Hz, (b) both Tx laser and LO laser linewidths are 500 kHz, (c) Tx laser linewidth is 0 Hz and LO laser linewidth is 1 MHz. w/o: without, w/: with.

The distribution of the converged tap weights in the LMS dispersion equalizer for the system with different combinations of Tx and LO laser linewidth (still a constant summation) is illustrated in Fig. 8, where the transmission fiber length is 40 km. We can find that the tap weights give a very similar distribution for different combinations of the Tx and the LO laser linewidth. This can also reflect that the phase noise from the Tx laser and the LO laser gives an equal contribution for the EEPN in the LMS equalization, since the Tx laser and the LO laser phase noise interacts equally with the tap weights in the LMS equalizer.

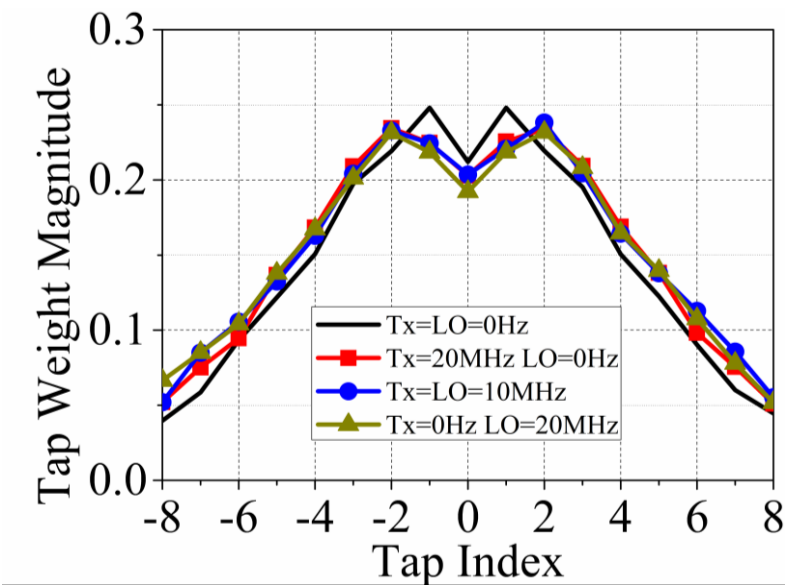

**Fig. 8.** Converged tap weights in LMS dispersion compensation algorithm with different distributions of Tx and LO laser linewidth.

Figure 9 shows the BER performance of the 28-Gbaud DP-QPSK transmission system using dispersion compensating fibers based ODC. It can be seen from Fig. 9(a) to Fig. 9(c) that the system gives the same performance with different combinations of Tx laser and LO laser linewidth, as well as different fiber dispersion, and no EEPN effects has been found. Figure 10 shows the performance of the transmission system using DCF and chirped FBG for dispersion compensation respectively, where the FBG based dispersion compensation gives an identical performance as the system using DCF based dispersion compensation. Therefore, there is no EEPN in the transmission system using optical dispersion compensation. However, in the transmission systems using ODC, the management and the mitigation of the fiber nonlinearities will become a critical issue, and this point will be discussed in more detail in Section 7. It is noted that here we assume an ideal optical dispersion compensation in the use of both DCF and FBG. In the case of DCF based compensation, the fiber dispersion is compensated using the DCF applied in the end of total optical link [2,44], where 100% of the dispersion is assumed to be compensated. In the FBG based compensation, the ideally channelized FBG is applied for the fiber dispersion compensation, where the effect of phase ripple is neglected [45,46]. In practical applications, the non-perfect compensation in the ODC could happen due to the inaccurate information of the fiber dispersion in the transmission link. The residual dispersion can be compensated electronically using an adaptive dispersion equalizer, and in this case small EEPN effect from the interaction between the laser phase noise and the adaptive dispersion equalizer has to be considered.

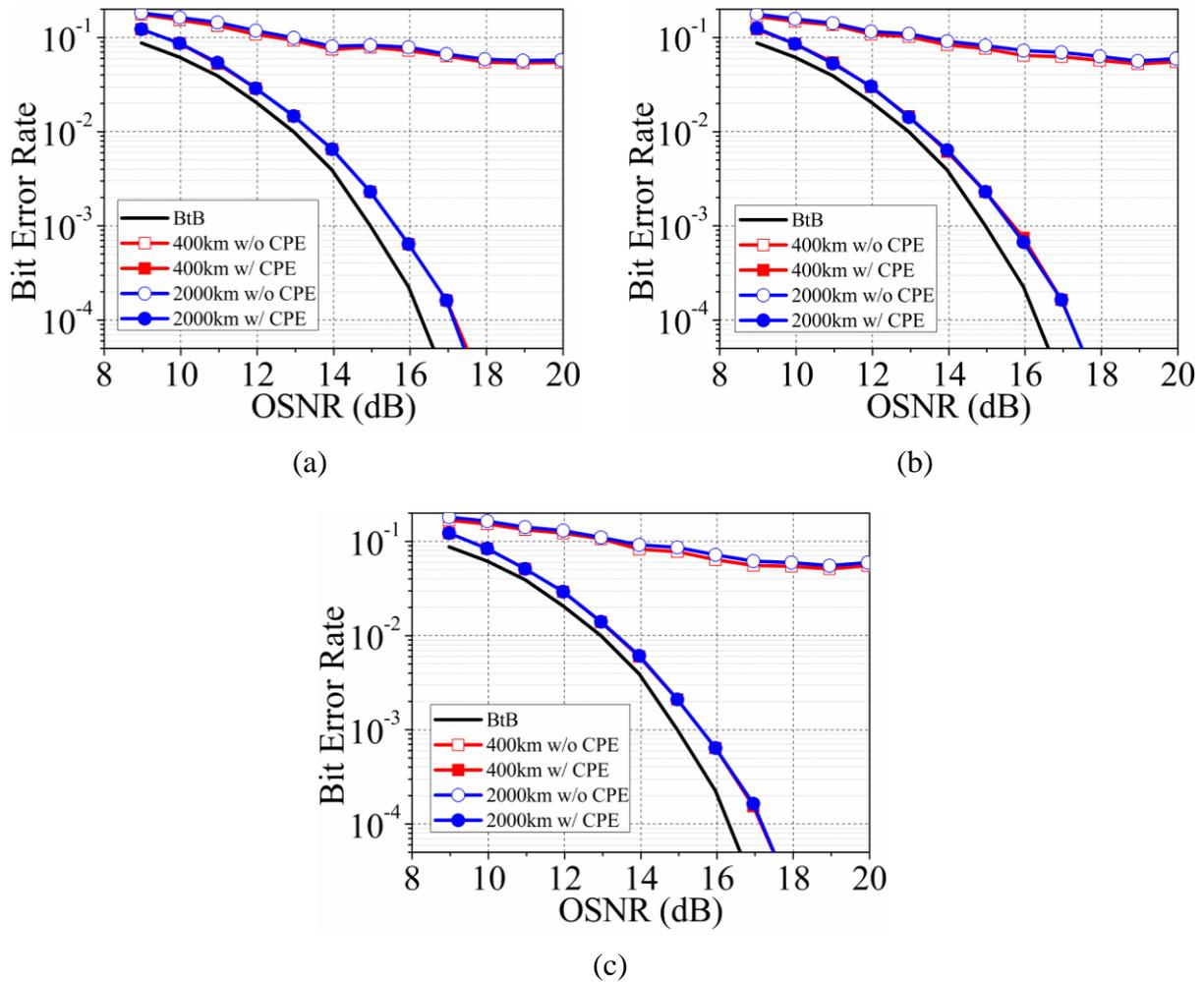

**Fig. 9.** Performance of 28-Gbaud DP-QPSK transmission system with different fiber lengths using DCF based optical CD compensation. (a) Tx laser linewidth is 4 MHz and LO laser linewidth is 0 Hz, (b) both Tx laser and LO laser linewidths are 2 MHz, (c) Tx laser linewidth is 0 Hz and LO laser linewidth is 4 MHz. w/o: without, w/: with.

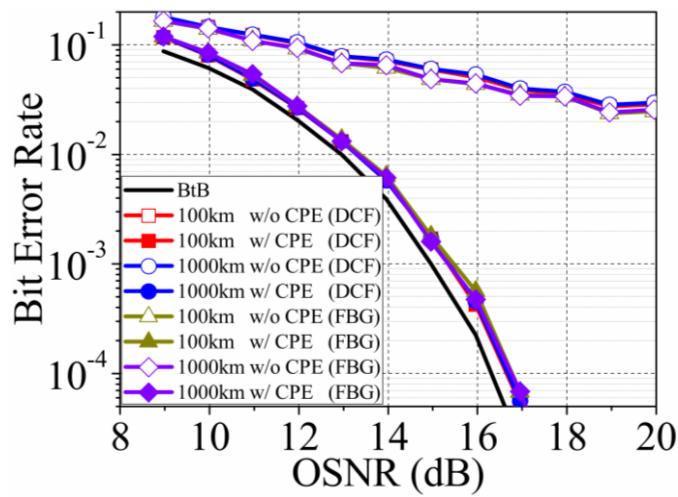

**Fig. 10.** Performance of 28-Gbaud DP-QPSK transmission system with different fiber lengths using optical CD compensation based on DCFs and chirped FBGs (both Tx and LO lasers linewidths are 500 kHz).

A comparison of the three electronic dispersion compensation methods (the TDE, the FDE, the LMS) in terms of the EEPN induced OSNR penalty at BER = $10^{-3}$ (compared to the BtB case) for different effective linewidths is shown in Fig. 11, where the one-tap NLMS algorithm is applied for CPR. The effective linewidth is obtained from Eq. (5), and the testbed is still based on the 28-Gbaud DP-QPSK coherent transmission system. It can be seen that the TDE and the FDE dispersion compensation show a similar performance, while the OSNR penalty in the LMS dispersion compensation grows much faster with the increment of the effective linewidth, due to a more severe EEPN generated in the LMS dispersion equalization.

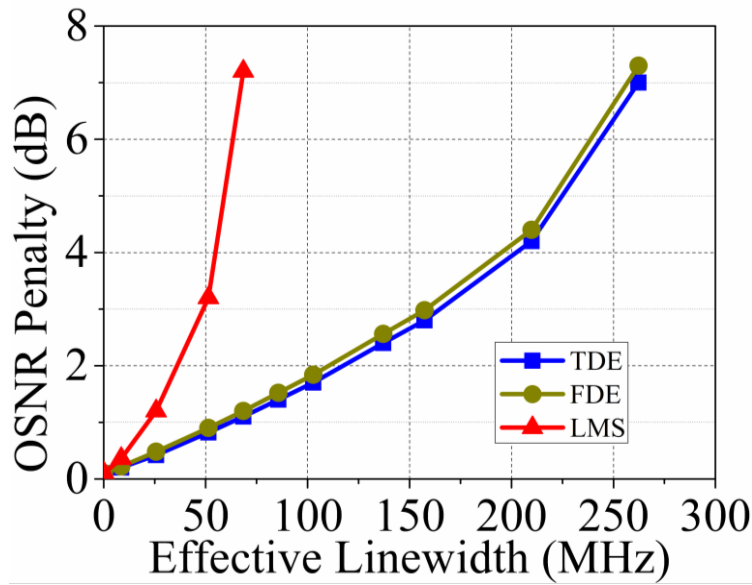

**Fig. 11.** The EEPN induced OSNR penalty (compared to BtB) at BER = $10^{-3}$ using different electronic CD compensation methods.

**6.2 Influence of carrier phase recovery on EEPN**

To study the EEPN effects in the transmission system using different carrier phase recovery approaches, we have investigated the EEPN induced BER floors in the 28-Gbaud DP-QPSK coherent optical transmission system, as shown in Fig. 12, where the three carrier phase recovery algorithms (the one-tap NLMS, the BWA, and the VV) are applied respectively. The optical fiber length of 2000 km and the FDE dispersion compensation are used in the simulation work. All the BER floors are measured at the ONSR value of 40 dB. The linewidths of the Tx and the LO lasers are both 5 MHz in Fig. 12(a), and the linewidths of the Tx and the LO lasers are both 10 MHz in Fig. 12(b). The numerical results are obtained using different block size in the BWA and the VV carrier phase recovery algorithms, and an optimum step size parameter in the one-tap NLMS algorithm. The trends of the results in Fig. 12(a) and Fig. 12(b) are identical. The block-wise average algorithm performs slightly better than the one-tap NLMS algorithm, when the block size is less than 11 in Fig. 12(a) and less than 9 in Fig. 12(b). The Viterbi-Viterbi algorithm works marginally better than the one-tap NLMS method, when the block size is less than 21 in Fig. 12(a) and less than 15 in Fig. 12(b). Meanwhile, the Viterbi-Viterbi algorithm shows a marginal improvement compared to the block-wise average algorithm, while it requires more computational complexity. It is noted that the effective linewidth for indicating the amount of EEPN includes the impacts from both the fiber dispersion and the laser linewidths, therefore, the results in Fig. 12 can also demonstrate the trend of the BER floors in different CPR algorithms under the use of different fiber dispersion.

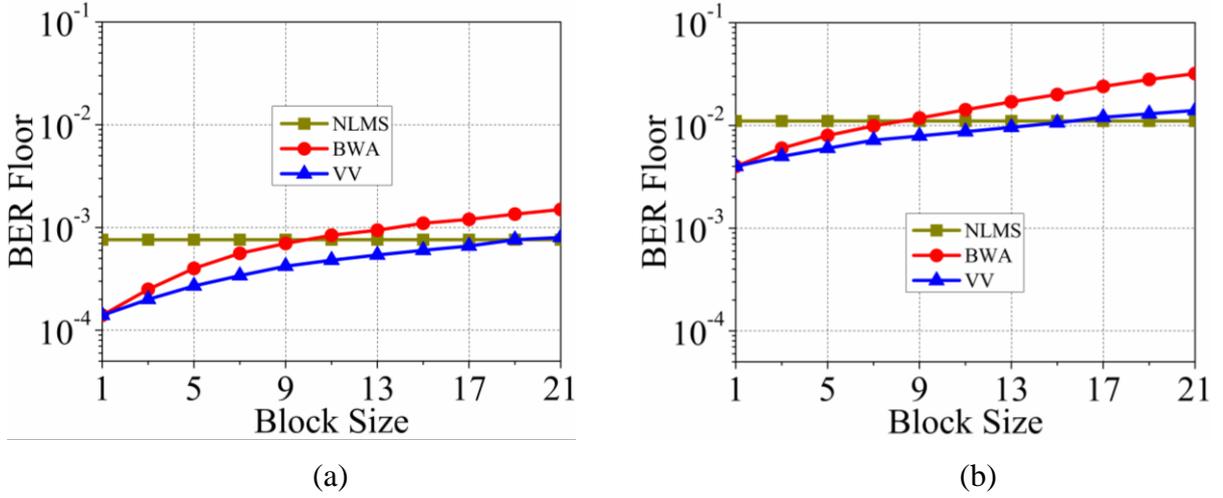

(a)                      (b)

**Fig. 12.** BER floors in 28-Gbaud DP-QPSK coherent transmission system using different CPR algorithms considering EEPN. (a) both Tx laser and LO laser linewidths are 5 MHz, (b) both Tx laser and LO laser linewidths are 10 MHz.

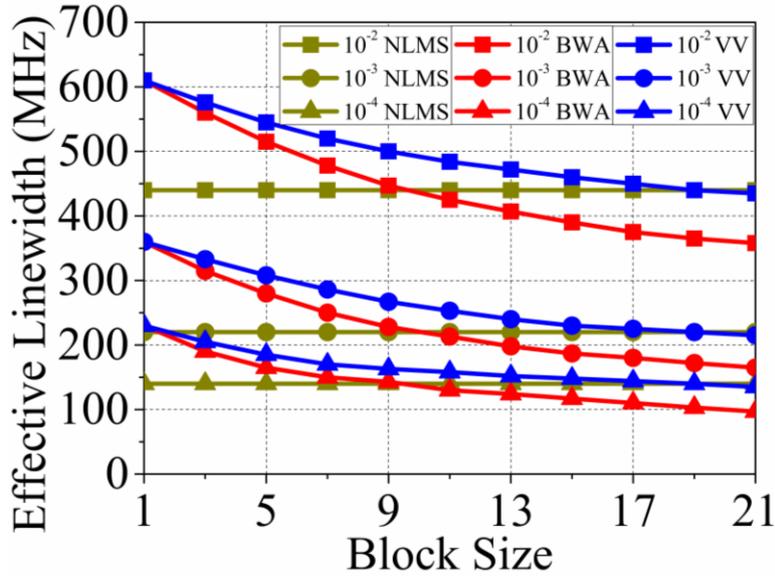

**Fig. 13.** Effective linewidth tolerance in the 28-Gbaud DP-QPSK transmission system at different BER floors ($10^{-2}$, $10^{-3}$, $10^{-4}$) for different CPR approaches.

The maximum tolerable effective linewidth in the 28-Gbaud DP-QPSK coherent transmission system is investigated in Fig. 13 (consistent with the analyses in Ref [32]), where the three carrier phase recovery algorithms (the BWA and the VV methods with different block size, the one-tap NLMS method with an optimum step size) are applied. Here the maximum tolerable effective linewidth refers to the effective linewidth in the communication system which leads to a certain BER floor ($10^{-2}$, $10^{-3}$, or $10^{-4}$) in the carrier phase recovery. We can find that the maximum tolerable effective linewidth in the block-wise average and the Viterbi-Viterbi CPR approaches degrade with the increment of the block size. The BWA algorithm performs better (allowing a larger effective linewidth) than the one-tap NLMS algorithm when the block size is less than 11, and the Viterbi-Viterbi algorithm performs better than the one-tap NLMS algorithm when the block size is less than 21. In accordance with previous results, the Viterbi-Viterbi algorithm does not show a considerable improvement in term of the tolerable effective linewidth either, compared to the block-wise average algorithm.

## 7. Discussions

According to the above descriptions, the EEPN does not exist in the coherent transmission systems using optical dispersion compensation, such as DCFs and chirped FBGs, and it seems the linear distortions can be less in such systems. However, the mitigation of the fiber nonlinear effects in the systems using ODC should be significantly considered. One feasible method can be the optical back-propagation (OBP), to compensate the chromatic dispersion and the nonlinearities simultaneously [44,47]. Thus both EEPN and fiber nonlinearities can be mitigated with a low complexity in the coherent transmission systems using OBP. Another promising option is the optical phase conjugation (OPC), which can cancel the fiber nonlinear effects by utilizing the phase conjugate of the transmitted signal to generate an opposite nonlinear phase shift [48,49]. The combination of the OPC and the ODC can also be a feasible approach to compensate the EEPN and the fiber nonlinearities at the same time [50].

All the numerical simulations in this paper have been carried out based on the 28-Gbaud NRZ-DP-QPSK coherent transmission system, which corresponds to a standard 100-Gbit/s optical fiber networks. As a matter of fact, the indications and the conclusions are also appropriate and can be transplanted for the $n$-PSK and the $n$-QAM coherent transmission systems, and the impacts of EEPN will be more significant and critical for such systems using a higher-order modulation format. The TDE and the FDE dispersion compensation methods are independent of modulation format, and the LMS dispersion equalization and the one-tap NLMS CPR algorithm can also be applied directly in the $n$-PSK and the $n$-QAM transmission systems. The block-wise average and the Viterbi-Viterbi algorithms can be used in the $n$-PSK and the star $n$-QAM systems directly. But for the square $n$-QAM transmission systems, some additional DSP operations, such as constellation partitioning and symbol classification, have to be accommodated according to the specific modulation format [51,52].

In the descriptions regarding CPR, the performance of the block-wise average and the Viterbi-Viterbi algorithms degrades with the increment of the block-size, since only the BER floors are employed for evaluating the transmission systems, where the additive channel noise, e.g. amplified spontaneous emission (ASE) noise, is neglected. Actually, a larger block size in the BWA and the VV algorithms is helpful for mitigating the additive channel noise, although it will degrade the tolerance of the CPR approach against the laser phase noise. As shown in Fig. 12 and Fig. 13, the performance dependence on the block size in the BWA and the VV algorithms is not strong, meaning that a large block size (up to ~21) can be employed to mitigate the additive white Gaussian noise (AWGN) in the practical coherent optical transmission networks. Meanwhile, the one-tap NLMS algorithm can also give a good accommodation for the additive channel noise in the transmission system, if a optimum step size is applied [9,17].

In addition, it is noted that the block-wise average and the Viterbi-Viterbi algorithms require the use of differential encoding to avoid the cycle slip effect in carrier phase recovery, while the one-tap NLMS algorithm does not require a differential encoding [9,18,19]. The use of differential encoding generally leads to an additional OSNR penalty, which also needs to be taken into consideration in practical applications of carrier phase recovery approaches. In this work, we employed the coherently-detected DP-QPSK in all the transmission scenarios without considering the impact of differential encoding, and in practical communication systems the cycle slip problem can be solved by using either pilot-symbol assisted estimation or forward error correction (FEC) coding techniques [53,54].

## 8. Conclusions

We have presented a comprehensive investigation on different chromatic dispersion compensation and carrier phase recovery approaches in the *n*-PSK and *n*-QAM coherent optical transmission systems, considering the impacts of EEPN. Four CD compensation approaches, including the time-domain equalization, the frequency-domain equalization, the least mean square adaptive equalization for EDC, and the dispersion compensating fiber for ODC, have been applied for dispersion compensation. Three carrier phase recovery methods are also employed for carrier phase estimation: the one-tap NLMS algorithm, the block-wise average algorithm, and the Viterbi-Viterbi algorithm. The impact and the origin of EEPN are analyzed and discussed in detail by using and comparing different dispersion compensation and carrier phase recovery approaches. Numerical simulations have been implemented in the 28-Gbaud NRZ-DP-QPSK coherent optical transmission system, and the results show that the origin of EEPN depends on the choice of digital dispersion compensation methods and the effects of EEPN behave moderately different in different carrier phase recovery approaches. In the transmission system employing the TDE and the FDE for dispersion equalization, the system performance is significantly impacted by the equalization enhanced LO phase noise. However, in the LMS adaptive dispersion equalization, the system performance is equally influenced by the equalization enhanced Tx phase noise and the equalization enhanced LO phase noise. Meanwhile, the LMS adaptive equalization is less tolerant against phase fluctuation than the TDE and the FDE, when the CPR is employed. There is no EEPN in the system using optical dispersion compensation. In the comparative study of carrier phase recovery, the one-tap NLMS algorithm performs slightly worse (still acceptable) than the block-wise average and the Viterbi-Viterbi algorithms. The Viterbi-Viterbi approach offers a marginal improvement compared to the block-wise average approach, while it requires more computational complexity.

Our analysis and discussions are useful and important for the practical application and design of the DSP modules in the long-haul high speed coherent optical fiber transmission systems, where the EEPN can not be neglected.

**Acknowledgements**

This work is supported in part by UK Engineering and Physical Sciences Research Council (UNLOC EP/J017582/1), in part by European Commission Research Council FP7-PEOPLE-2012-IAPP (GRIFFON, No. 324391), and in part by Swedish Research Council Vetenskapsradet (No. 0379801).